\documentclass[12pt]{aastex}
\usepackage{emulateapj5}
\usepackage{apjfonts}
\usepackage{epsf}
\usepackage{graphicx}


\lefthead{INOUE and CHIBA}
\righthead{Direct Mapping of Massive Compact Objects in Extragalactic Dark Halos}

\newcommand{\f}{\frac}

\newcommand{\bb}{\bibitem}
\newcommand{\BF}{\begin{figure}\begin{center}}
\newcommand{\EF}{\end{center}\end{figure}}
\newcommand{\BE}{\begin{equation}}
\newcommand{\EE}{\end{equation}}
\newcommand{\BEA}{\begin{eqnarray}}
\newcommand{\EEA}{\end{eqnarray}}
\newcommand{\ti}{\textit}

\newcommand{\ms}{M_{\odot}}
\begin{document}

\submitted{\rm\it NAOJ-Th-Ap2003, No.22}

\title{Direct Mapping of Massive Compact Objects 
in Extragalactic Dark Halos}
\author{Kaiki Taro Inoue and Masashi Chiba}

\affil{Division of Theoretical Astrophysics,
National Astronomical Observatory,
2-21-1 Osawa, Mitaka, Tokyo 181-8588, Japan;
\email{tinoue@th.nao.ac.jp, chibams@gala.mtk.nao.ac.jp}}
\date{\today}

\maketitle
\begin{abstract}
A significant fraction of non-baryonic or 
baryonic dark matter in galactic halos
may consist of MASsive Compact Objects (MASCOs) 
with mass $M=10^{1-4}M_{\odot}$.
Possible candidates for such compact objects include primordial black holes or
remnants of primordial (Population III) stars.
We propose a method for directly detecting MASCOs
in extragalactic halos, using the VLBI techniques with extremely 
high resolution that would be achieved by the 
next generation mission of the VLBI Space 
Observatory Program such as VSOP-2. If a galactic halo comprising 
a large number of MASCOs produces multiple images of a 
background radio-loud QSO by
gravitational lensing, then a high-resolution radio map of each macro-lensed
image should reveal microlensing effects by MASCOs.
To assess their observational feasibility, we simulate microlensing of the
radio-loud, four-image lensed QSO, B1422+231, assuming angular resolution of
$\!\sim\!$0.01 mas. MASCOs are represented by point masses. 
For comparison, we also simulate 
microlensing of B1422+231 by singular isothermal spheres.
We find that the surface brightness of
the macro-lensed images shows distinct spatial patterns on the scale of
the Einstein radius of the perturbers.
In the case of point-mass perturbers, many tiny dark spots also appear 
in the macro-lensed images associated with a decrease
in the surface brightness toward the fringe of the original QSO image, whereas
no such spots are available in the SIS models.
Because such spatial patterns in each macro-lensed image 
cannot be linearly mapped to those in other macro-lensed images 
if they are relevant to lensing perturbers, it is
fairly easy to discriminate them from intrinsic substructures within a QSO.
Based on the size, position and magnified or demagnified patterns of images,
we shall be able to determine the mass and density profile of an individual
MASCO as well as its spatial distribution and abundance in a galactic halo.

\end{abstract}


\section{introduction}
Gravitational lensing is a powerful tool for probing dark matter in the form
of compact objects, such as black holes, neutron stars, and dwarfs.
To date, various kinds of gravitational lensing studies give stringent upper
limits on the abundance of these objects on very different mass scales,
ranging from substellar objects $\sim 10^{-7}\ms$ (Alcock et al. 2000)
to galaxy clusters $\sim 10^{14}\ms$ (Nemiroff 1991). 

However, there have been no stringent observational constraints on dark compact
objects with a mass range of $10^{1}\ms \lesssim M \lesssim 10^{4}\ms$
corresponding to the mass range of massive stars. For brevity, we call them
MASCOs (MASsive Compact Objects). 
It is of particular importance to constrain their abundance in galactic
dark halos, because MASCOs are both baryonic and non-baryonic dark matter
candidates. Recent observation of the cosmic microwave background (Bennett
et al. 2003) suggests that the primordial stars (or Population III stars) are rather
massive as indicated by theoretical works (Bromm, Coppi, \& Larson 1999; Abel,
Bryan, \& Norman 2000; Tsuribe \& Inutsuka 2001; Omukai \& Palla 2002) and 
their formation efficiency are higher than current theoretical works suggest
(Cen 2003). In some MASCO mass scales, primordial stars collapse directly
to form black holes (Bond et al. 1984; Woosley 1986). Thus, a large
fraction of baryonic dark matter in galactic 
dark halos may consist of black holes
that are remnants of massive primordial stars.  
Alternatively, as the cold dark matter (CDM) candidates, 
MASCOs may be primordial black holes
that are produced in the early universe at the era between the QCD phase
transition and the big-bang nucleosynthesis (Carr 1975).

To date, various gravitational lensing methods that can probe
MASCOs have been proposed. Gould (1992) proposed that 
annual oscillation in the light curve of a MACHO lensing event can 
distinguish a MASCO from a stellar mass compact object. However, for a mass
range of $10^{2}\ms \lesssim M   \lesssim  10^{3}\ms$, the annual
oscillation owing to the Earth's motion is too small to determine
the transverse velocity. Totani (2003) argued that observation of 
cluster-cluster microlensing system by a 8~m-class
ground-based telescope can probe the abundance 
of compact objects with mass $10^{-5}\ms \lesssim M  \lesssim  10^{12}\ms$.  
Although his proposed method covers a wide range of mass scales, 
the method itself cannot break the degeneracy in the lensing parameters 
such as mass or transverse velocity 
owing to the lack of information of the source star luminosity.

As an alternative, we propose a method to directly detect MASCOs 
using the VLBI techniques with extremely high resolution. 
Suppose that a distant radio-loud QSO shows multiple images owing to
the lensing effect of a foreground galaxy. 
If there are a large number of MASCOs in the galactic halo,
then the high-resolution radio map of the 
gravitationally lensed QSO images should reveal the microlensing
effects by MASCOs. In contrast to the previous methods,
our proposed method can determine the mass of MASCOs in the dark
halo without any assumption on the lensing parameters.  
Furthermore, we can also prove the spatial distribution of
MASCOs in the dark halo projected onto the surface perpendicular to the
line of sight. A possible target lens system is  
B1422$+$231 in which the source is a radio-loud QSO at 
redshift $z_s=3.62$ and the lens is 
an elliptical galaxy at redshift $z_L=0.34$. 
Although the method we propose here is similar to that of Wambsganss
\& Paczy'nski (1992) in which the target compact object mass is
set to $M\sim 10^{6}\ms$, the lensing effect by MASCOs can be quite different
since the Einstein radius is much smaller than the size of a source and the
total number in a dark halo is large. 
In this \ti{Letter}, we show our simulation results for
the lens system B1422$+$231 microlensed by MASCOs, that
would be observable by the future VLBI network. 

We propose to construct a new VLBI network with extremely high
resolution such as the planned next generation VLBI Space Observatory Program
(VSOP-2) (Hirabayashi et al. 2001) and produce 
high resolution maps of strongly lensed radio-loud QSOs
such as B1422+231. Such maps might provide us with valuable information of 
the first star remnants and the initial condition of density fluctuations on 
MASCO scales as well as the nature of the CDM.

\section{Method}

We consider a lensing system in which a radio-loud QSO is strongly
lensed by the dark halo of an intervening galaxy.
If MASCOs constitute a large amount of the dark matter component, then we
expect to see many microlensing events by them, because the microlensing
optical depth is nearly 1 in a strong lens.
In order to detect the signatures of such microlensing in the
macro-lensed QSO images, one needs extremely high-resolution radio
mapping of the images.
The necessary resolution can be estimated by the Einstein radius of MASCOs 
\BE
\theta_E\sim3\times10^{-2} \Biggl ( \f{M}{10^2 M_{\odot}}
\Biggr)^{\f{1}{2}} \Biggl ( \f{D_L D_S /D_{LS}}{\textrm{Gpc}}
\Biggr)^{-\f{1}{2}} \textrm{mas},
\EE
where $D_L$, $D_S$, and $D_{LS}$ are the angular diameter distance to
the lens, source, and distance to the lens from the source, respectively.  
Although current observations cannot achieve such a high-resolution, it 
is not out of reach in the near future. 
For instance, the target resolution of the VSOP-2 mission is 
$\sim 0.025$ mas at 43 GHz, which can resolve MASCOs with 
mass $M\gtrsim 10^{2}\ms$ for lensing systems at
1~Gpc. The VSOP-2 satellite will be placed at an elliptical orbit
with an apogee height of $\sim$30,000 km and a perigee height of
$\sim$1,000 km (Hirabayashi et al. 2001). The resolution can be further 
improved by a factor of two or three by placing satellite(s) at a 
much distant orbit or by increasing the observational frequency.  

In order to demonstrate the microlensing effects of MASCOs,
we consider the QSO lensing system B1422+231 whose images consist of three 
highly magnified ones A, B, and C, and a faint one D located near the 
lens galaxy. The redshifts of the source and the lens are
$z_S=3.62$ and $z_L=0.34$, respectively. In what follows, we assume 
the following cosmological parameters: $\Omega_m=0.3$,
$\Omega_\Lambda=0.7$ and $h=0.7$, giving $D_L=1.00$ Gpc and $D_S=1.49$ Gpc. 

To model the macro-lensing system, we adopt
a singular isothermal ellipsoid (SIE) in an external shear field in which the
isopotential curves in the projected surface perpendicular to the
line of sight are ellipse (Kormann et al. 1994).
This lens model has the following parameters: 
Einstein angular radius $\theta_E$, a ratio $q$ of the minor axis to the
major axis of the ellipse of the iso-potential contour, position angle $\phi_g$  
of the minor axis, source position on the source plane, lens position on
the lens plane, unlensed source flux, strength and direction of the external
shear $(\gamma,\phi_s)$. The best-fit lensing parameters using $\chi^2$ fit is 
$\theta_E=0.78$ arcsec, $q=0.84$, $\phi_g=-57.5^\circ$, $\gamma=0.21$, 
$\phi_s=-53.9^\circ$ (Chiba 2002), where the one-dimensional velocity dispersion
of the lens is estimated as $\sigma=183.23$ km~s$^{-1}$.
The observed position of each image is taken from HST observations with
the Faint Object Camera by Impey et al (1996). We show, in the middle panel of
Figure 1, the model results for the four lensed images.
It should be noted that B1422+231 shows an anomalous flux
ratio in the macro-lensed images possibly caused by the CDM subhalos with masses
larger than $10^6 M_\odot$ (Chiba 2002). We neglect below the effects of such
subhalos which are not relevant to microlensing by MASCOs. 

We estimate the surface number density of MASCOs with an individual
mass of $M$ in the lens galaxy. Because the ellipticity of the lens in
B1422+231 is modest, it can be approximated by that of a singular isothermal
sphere (SIS) with surface mass density $\Sigma(R)=\sigma^2/2 G R$ where
$R$ is the projected radial distance from the lens center.
The radial distances of the images A, B, and C from the center are roughly
equal to the linear Einstein radius, $\xi_0 \equiv D_L \theta_E = 3.8$ kpc.
Note that 1~mas corresponds to $\!\sim\!4.9$~pc on the lens
plane. Then the surface number density of MASCOs at $R = \xi_0$ is estimated as 
$N(\xi_0)=f \Sigma(\xi_0)/M\sim f \times 2.4\times 10^2 
(M/10^2\ms)^{-1}$~mas$^{-2}$,
where $f \le 1$ is a parameter that represents the fraction of
MASCOs in the total mass of the dark halo. 
For the D image, the surface number density will be 2.7 times larger because 
the radial distance is as small as $R=\xi_0/2.7$.
Assuming that the observable radio-emitting region of the QSO has a typical
linear size of the order of $\sim 10$~pc, then the total number of MASCOs
overlaid on the macro-lensed images would be several for $M=10^2\ms$ even if the 
MASCO fraction is of the order of a few percents. 
The presence of more extended jets in the images will increase this number.

First, we assume that MASCOs are placed randomly on the lens plane
in which the mean number is determined by the surface number density $N(R)$.
Clustering of MASCOs on subgalactic scales will not be considered for the sake
of simplicity; if a clustering signature is observed in the lensed images,
it will give us a valuable information about the formation process of MASCOs.

Second, we calculate the total gravitational potential of
the lens from the sum of a SIE that best fits the position and the flux
ratios of the four macro-lensed
images and perturbers, whereas the consistent
calculation should arrange this combined lens potential rather than the
potential of the SIE alone.
However, this simplification does not affect the concerned microlensing
signatures by MASCOs.
In our simulation, MASCOs are represented by point masses which perturb
the macro-lens. We also carry out simular calculation in which the perturbers are 
represented by SISs for comparison.
We also assume that the unlensed QSO image has a circular symmetry with
a radial profile given by the Gaussian distribution with a standard deviation $3h^{-1}$~pc.

\begin{figure*}
\centerline{\includegraphics[width=17cm]{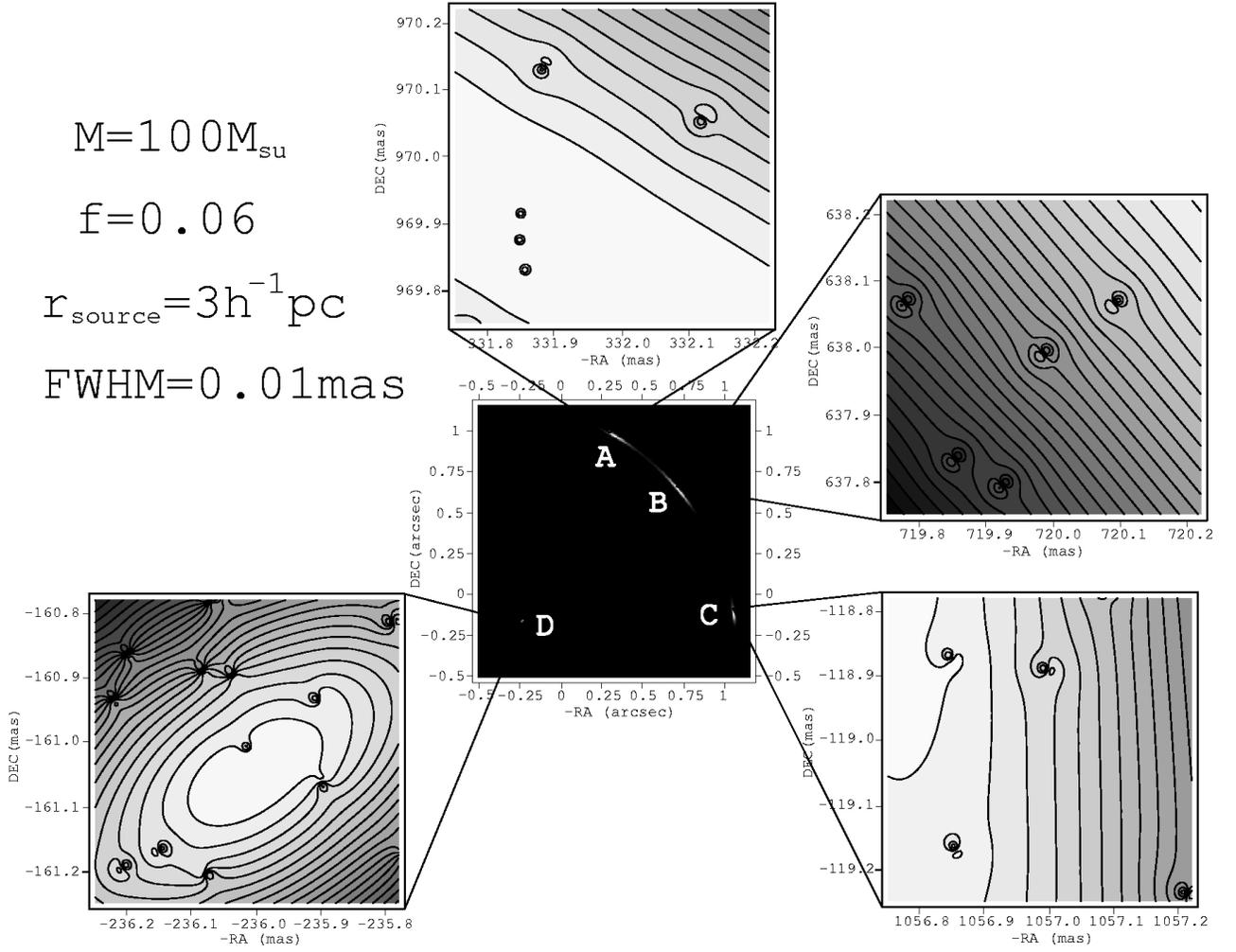}}
\caption{A simulation of radio maps of B1422+231
smoothed by a Gaussian beam with FWHM$\!=\!0.01mas$. MASCOs are represented
by point masses with mass $M=10^2\ms$.
Dark spots in zoomed four images correspond to
MASCOs in the dark halo of the foreground galaxy. The 
iso-surface brightness contour interval is 1/40 of the maximum
 value. The MASCO fraction to the dark matter halo is assumed to be
 $f\!=\!0.06$ corresponding to MASCO density $\Omega_\ast\!\sim\!0.018$ provided that
 the total matter density is $\Omega_m\!=\!0.3$.}
\end{figure*}

\begin{figure*}
\centerline{\includegraphics[width=6cm]{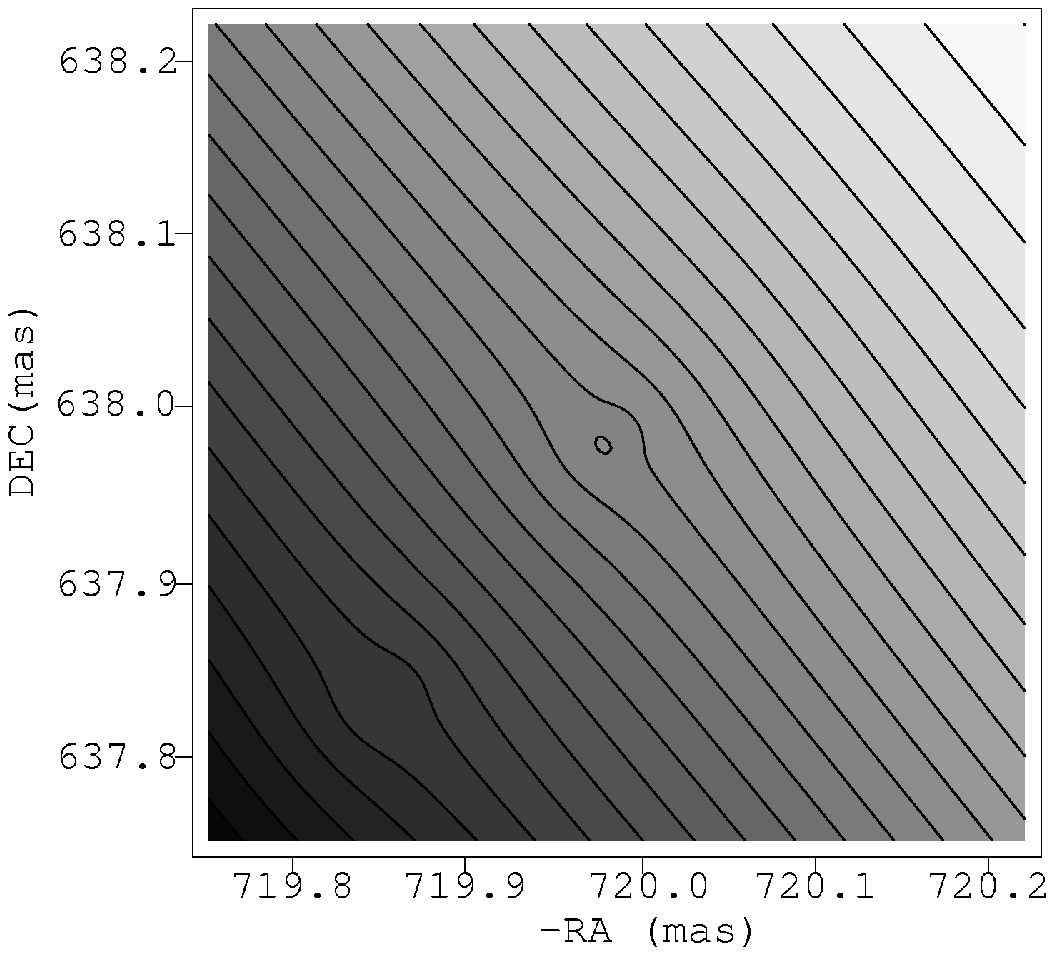}
            \includegraphics[width=6cm]{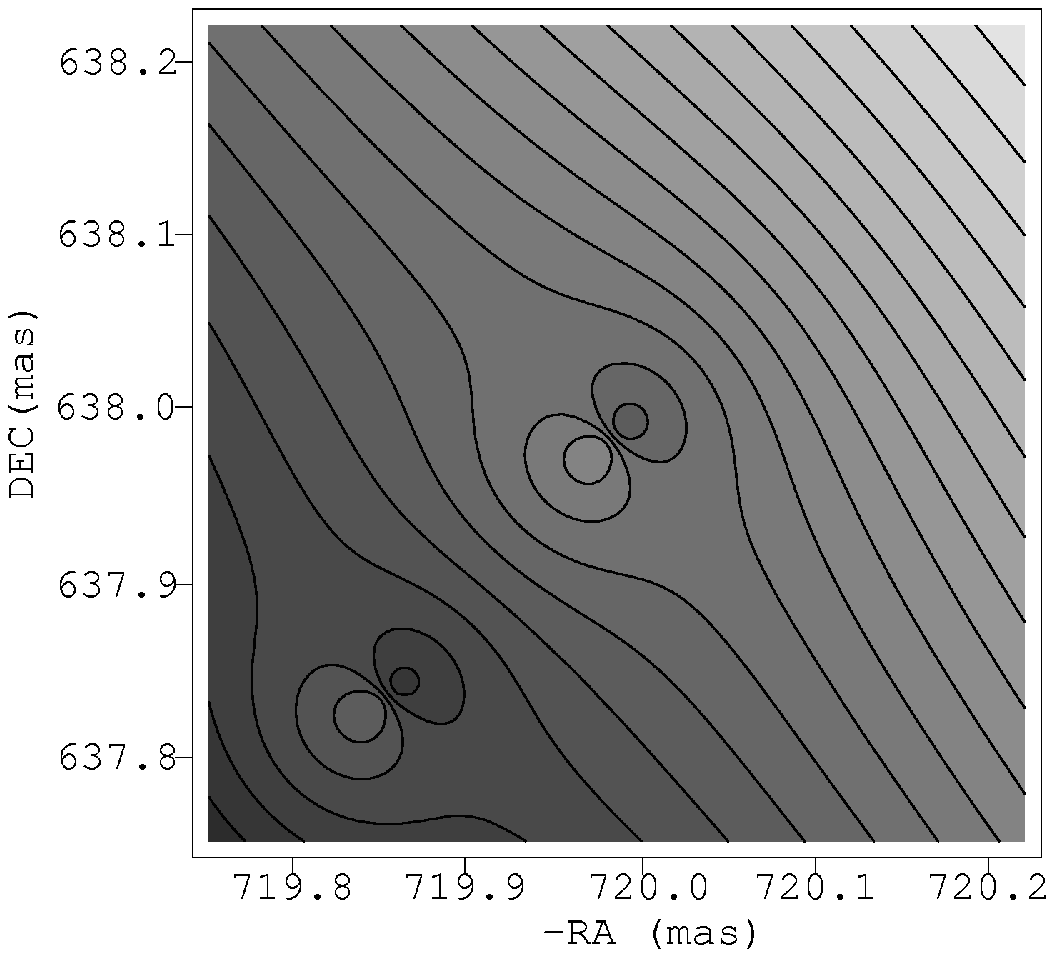}}
\caption{Simulations of radio maps of B image in B1422+231
smoothed by a Gaussian beam with FWHM$\!=\!0.01mas$ in which singular 
isothermal spheres with mass $M=10^4\ms$ are placed randomly 
in the dark halo. The ratio of the SIS
radius to the tidal disruption radius is 1(left) and 1/4(right). }
\end{figure*}

\section{Results}
Four panels in Figure 1 show the portions of four macro images disturbed by
the point-mass MASCOs with $M=10^2 M_\odot$ and $f=0.06$, where a resolution of
0.01~mas is assumed. One can clearly see several dark spots in the brightest
part of the macro images as well as several neighboring pairs of dark and bright
spots in the region where the surface brightness of the image has a spatial
gradient. These characteristic properties of microlensing by point
masses can be explained in the following manner.

Let a point mass be located at $\textbf{x}=0$ on the lens plane for the
extended image on the source plane with coordinate $\textbf{y}$, where both
$\textbf{x}$ and $\textbf{y}$ are measured in units of the Einstein radius
of a MASCO. The lens equation is then given by
$\textbf{y}=\textbf{x}-\textbf{x}/x^2$. It follows that
the positions on the lens plane within the Einstein radius satisfying
$-0.62 \lesssim x <0$ are mapped to $y>1$. Conversely, the source image
extended outside the Einstein radius is mapped to the lens-plane region
with radius $x=0.62$ inside the Einstein radius. Suppose that the
surface brightness of the source
has a profile $\sigma_y=y^{-\beta}$ with $\beta>0$. Then, in the neighborhood
of the lens center $x=\epsilon \ll 1$ with $\epsilon>0$, the surface brightness
of the lensed image, $\sigma_x$, is decreased as $\sigma_x=\epsilon^\beta$.
Thus, if $\sigma_y$ decreases monotonically with increasing distance
from the lens center, then one would observe a dark spot (inside the Einstein
radius) at the lens center.
If the sign of $\beta$ is minus, then one would observe
a bright spot within the Einstein radius. Therefore, there will be
a neighboring pair of dark and bright spots (within the Einstein radius)
in the region mapped from the source plane where the gradient of $\sigma_y$
has a non-vanishing constant value. For less massive MASCOs,
such discontinuous changes of the lensed image become inconspicuous
because the change in the surface brightness on the scale of the Einstein
radius becomes smaller. However, if the source has an intrinsic substructure,
it may have a locally large gradient in its surface brightness, yielding pairs
of dark and bright spots in the lensed image even for less massive MASCOs.
It is worth noting that we can easily distinguish these microlensing effects
from any intrinsic structures of the source by linearly mapping
the macro-lensed image to one of the other images. 

Figure 2 shows the microlensing effects of the SISs with $M=10^4 M_\odot$ on the
B image. In this case, the size of each SIS, $r_{SIS}$, is a free parameter,
affecting the strength of the lensing effect. Its upper limit is set by a tidal
radius, $r_t = (GMr^2 / \pi \sigma^2)^{1/3}$, at its position $r$ from the
galaxy center. We examine two cases, $r_{SIS}/r_t=1$ and $1/4$,
when $r$ is defined as the projected distance of the image B from the galaxy center,
i.e., $r=\xi_0$.
In the SIS model, the lens equation is given as
$\textbf{y}=\textbf{x}-\textbf{x}/x$.
In sharp contrast to the point mass model,
the source image extended outside the Einstein radius $y>1$ cannot be mapped to the region inside
the Einstein radius $x<1$, indicating that there will be almost
no visible effect from a lens placed in the region where the
gradient in the surface brightness within the Einstein radius vanishes.
On the other hand, in the region where the surface brightness gradient does not
vanish, a discontinuous change of the image brightness on the scale of the
Einstein radius is observed, as found for the point-mass model.
However, magnification or demagnification effect by SISs is less conspicuous
than the point-mass model.

\section{Summary and Discussion}
In this \ti{Letter}, we proposed a method to directly ditect MASCOs,
i.e., massive compact objects with mass $M=10^{1-4}M_{\odot}$ in 
extragalactic dark halos.
Using the VLBI techniques, radio maps of strongly
lensed radio-loud QSO with extremely high resolution $\sim 0.01$ mas
would reveal the abundance, the spatial distribution, and the mass density
profile of the MASCOs. 

We present numerical simulations for B1422+231, a typical strongly
lensed QSO system. In the presence of MASCOs, tiny dark spots and 
pairs of dark and bright spots will appear in the
macro-lensed images. From the size of the spot, one can estimate
the mass of the MASCO. 
Even if the total mass of MASCOs in the
dark halo of the lens galaxy constitutes only a few percents of the total
dark halo mass, we can detect MASCOs from characteristic substructures
in the map of the lensed QSO image
which cannot be linearly mapped to other macro-lensed images. 
If a perturbing object of the macro-lensed image is described by
a SIS, the signature of microlensing appears only if the gradient in
the surface brightness is very large or the size of the SIS is
sufficiently compact. In contrast to the point-mass
case, light from the fringe of the source is not visible within the
Einstein radius. Therefore, one can distinguish SISs from
point masses by analyzing the detailed spatial variation in the surface brightness.

If the WMAP result of the Thompson optical depth withstands future
data, we will have strong observational evidence that the primordial
stars are massive and their formation rate is quite large (Cen 2003).
For mass ranges of $50\ms \lesssim M \lesssim 100\ms$ 
and  $260 \ms \lesssim M$, the progenitor of the primordial gas
can directly collapse to form massive black holes without ejecting any metals
(Woosley 1986). Thus the primordial star
remnants may constitute a large fraction of the baryonic dark matter.
Because our proposed method is so powerful in constraining the
abundance of point-mass MASCOs, we might be able to constrain or determine  
the initial mass function of the primordial stars from the 
observed mass and the abundance of the MASCOs.

If the CDM is strongly clustered on MASCO mass scales, 
their subhalos might be detected by our proposed method if they are 
compact enough. In our simulation, with resolution of $0.01$ mas, 
SIS with mass $M \sim 10^4 M_\odot$ can be detected if the 
ratio of the tidal disruption radius to the SIS cutoff radius is $\lesssim 4$.  
By numerically estimating the abundance of subhalos on MASCO
scales, one can constrain the amplitude of the initial density fluctuations   
on such small scales from future radio observations.

\acknowledgments
We thank S. Kameno for useful comments on the specification of VSOP-2.  
This work was supported in part by Grant-in-Aid for 
Scientific Research Fund (No.11367).

\end{document}